\def\figcond{1}     
\def\figa#1#2#3{%
\begin{figure}[#2]
\ifnum\figcond>0
 \centerline{\hbox{\hsize=4.875in\hss\vbox to 3.45in{\vss%
 \vskip -0.15in%
 \centerline{\hskip 0.225in\epsfig{file=#1.ps,width=5.7in}}
 \vss}\hss}}
\else
 \centerline{\hbox{\hsize=4.875in\hss\vbox to 3.45in{\vss%
 \centerline{Figures or Hardcopys available from %
             koepf@josiah.tau.ac.il}%
 \vss}\hss}}
\fi
\caption{#3}
\end{figure}}
\def\figb#1#2#3#4{%
\begin{figure}[#3]
\ifnum\figcond>0
 \centerline{\hbox{\hsize=4.875in\hss\vbox to 3.45in{\vss%
 \vskip -0.15in%
 \centerline{\hskip 0.225in\epsfig{file=#1.ps,width=5.7in}}%
 \vss}\hss}}%
 \vskip 0.10in%
 \centerline{\hbox{\hsize=4.875in\hss\vbox to 3.45in{\vss%
 \vskip -0.15in%
 \centerline{\hskip 0.225in\epsfig{file=#2.ps,width=5.7in}}%
 \vss}\hss}}%
\else
 \centerline{\hbox{\hsize=4.875in\hss\vbox to 7.00in{\vss%
 \centerline{Figures or Hardcopys available from %
             koepf@josiah.tau.ac.il}%
 \vss}\hss}}
\fi
\caption{#4}
\end{figure}}
\ifnum\figcond>0
\documentstyle[preprint,floats,prc,aps,epsfig]{revtex}
\else
\documentstyle[preprint,floats,prc,aps]{revtex}
\fi
\begin{document}
\preprint{\hbox{\vbox{\hbox{DOE/ER/40427-12-N95}
         \vskip -.10in\hbox{nucl-th/9508048}}}}
\draft
\title{THE NUCLEON-NUCLEON INTERACTION IN THE CHROMO-DIELECTRIC
       SOLITON MODEL: DYNAMICS}
\author{S. Pepin and Fl. Stancu}
\address{Universit\'{e} de Li\`ege\\
         Institut de Physique B.5\\
         Sart-Tilman, B-4000 Li\`ege 1, Belgium}
\author{W. Koepf}
\address{School of Physics and Astronomy\\
         Raymond and Beverly Sackler Faculty of Exact Sciences\\
         Tel Aviv University, 69978 Tel Aviv, Israel}
\author{L. Wilets}
\address{Department of Physics, FM-15\\
         University of Washington, Seattle, WA 98195, USA}
\date{August 31, 1995}
\maketitle
\begin{abstract}

The present work is an extension of a previous study of the
nucleon-nucleon interaction based on the chromo-dielectric soliton
model. The former approach was static, leading to an adiabatic
potential. Here  we perform a dynamical study in the framework of the
Generator Coordinate Method. In practice, we derive an approximate
Hill-Wheeler differential equation and obtain a local nucleon-nucleon
potential as a function of a mean generator coordinate. This
coordinate is related to an effective separation distance between the
two nucleons by a Fujiwara transformation. This latter relationship is
especially useful in studying the quark substructure of light nuclei.
We investigate the explicit contribution of the one-gluon exchange
part of the six-quark Hamiltonian to the nucleon-nucleon potential,
and we find that the dynamics are responsible for a significant part
of the short-range N-N repulsion.

\end{abstract}
\bigskip
\section{Introduction}

In a previous investigation \cite{KW94}, we studied the N-N
interaction in the
framework of the chromo-dielectric soliton model from a static point
of view: we used the Born-Oppenheimer approximation to derive an
adiabatic N-N potential, which showed a soft core repulsion due
essentially to the color-electrostatic part of the one-gluon exchange.
Previous studies of the N-N interaction in terms of quark degrees of
freedom \cite{Fa83} have pointed out the importance of dynamical
methods (such as Generator Coordinate or Resonating Group) in the
calculation of a realistic N-N potential. For example, in a preceding
application of the non-topological soliton model to the N-N problem,
Schuh et al. \cite{Sc86} showed that a significant part of the
repulsion was due to dynamics; the absence of a repulsive core in some
previous works was also interpreted as an artifact of the adiabatic
approximation \cite{Fa83}.

The Lagrangian of the chromo-dielectric model is defined as in Ref.
\cite{KW94}:
\begin{equation}
{\cal L}={\cal L}_q+{\cal L}_\sigma+{\cal L}_G \ ,
\label{lag}
\end {equation}
with
\begin{eqnarray}
{\cal L}_q & = & \bar{\psi}\left(i\gamma^\mu D_\mu
                     - m_q\right)\psi \ , \\
{\cal L}_\sigma & = & \frac{1}{2}\partial{_\mu}\sigma\partial{^\mu}
                      \sigma-U\left(\sigma\right) \ , \\
{\cal L}_G & = & -\frac{1}{4} \kappa(\sigma) F^a_{\mu \nu}
                 F^{\mu\nu}_a \ ,
\end{eqnarray}
where $\psi$ is the quark operator and $m_q$ the current quark mass
matrix, set here to $m_q = 0$. The quark Lagrangian ${\cal L}_q$ is
expressed in terms of the covariant derivative
$D_\mu=\partial_\mu -ig_s T^a A_\mu^a$, and
$F^a_{\mu \nu}=\partial_\mu A_\nu^a-\partial_\nu A_\mu^a +
g_s f^{abc} A_\mu^b A_\nu^c$ is the $SU(3)$-color
tensor, where $f^{abc}$ are the $SU(3)$ structure constants and $T^a$
the $SU(3)$ generators. The quantity $U(\sigma)$ is the
self-interaction of the scalar field, $\sigma$, taken to be of
the form:
\begin{equation}
U(\sigma) = \frac{a}{2!}\sigma^2+\frac{b}{3!}\sigma^3+\frac{c}{4!}
            \sigma^4+B \ ,
\end{equation}
and the dielectric function $\kappa(\sigma)$ is:
\begin{equation}
\kappa(\sigma)= 1+\theta(\sigma)\left(\frac{\sigma}{\sigma_v}
\right)^{\!2}\left[2\frac{\sigma}{\sigma_v}-3\right] \ ,
\end{equation}
where $\sigma_v$ is the scalar field's vacuum expectation value and
$\theta$ the usual step function.

The quark self-energy, due to interactions with confined gluons in the
dielectric medium, generates an effective coupling
between the quarks and the scalar field:
\begin{equation}
{\cal L}_{q\sigma} = -g_{eff}(\sigma)\bar{\psi}\psi \ ;
\end{equation}
we choose $g_{eff}(\sigma)$ to be of the form:
\begin{equation}
g_{eff}(\sigma)=g_0\sigma_v\left(\frac{1}{\kappa(\sigma)}-1\right) \ .
\label{cou}
\end{equation}
The expression in Eq. (\ref{cou}) is an approximation to what has been
calculated in Ref. \cite{Kr88}, and it is constructed to simulate
spatial confinement already at the mean field level. Note that the
coupling in Eq. (\ref{cou}) breaks the chiral invariance of the
Lagrangian of Eq. (\ref{lag}). This is an example of dynamical
symmetry breaking from which a massless Goldstone boson emerges
naturally.

The parameters involved in our calculation are $a,\ b,\ c,\ g_0$ and
$\alpha_s=g_s^2/4\pi$,
as discussed in detail in Ref. \cite{KW94}. By fitting the
nucleon and the $\Delta$ masses and the proton's rms charge radius
one remains with two free parameters, for which it is convenient to
use the dimensionless quantities c and $f=b^2/ac$. In this paper, we
have chosen the set $f=\infty$ and c=10000 taken from Table 1 of Ref.
\cite{KW94}. Contrary to Ref. \cite{Sc86}, the quarks here are not
only coupled to the $\sigma$-field but also interact among themselves
through one-gluon exchange (OGE). The OGE is treated in Abelian
approximation, and it can be separated into two terms: a
self-interaction term (in addition to $g_{eff}(\sigma)$ of Eq.
(\ref{cou})), which is required for color confinement and which
contributes to the one-body part of the Hamiltonian, and a term of
mutual interactions, which gives rise to the two-body part of the
Hamiltonian. As mentioned earlier, in the adiabatic approximation of
Ref. \cite{KW94}, it was the color-electrostatic part of the OGE,
which arises from the time-component of the gluonic quadrivector
$A_\mu^c$, and especially the corresponding self-energy diagrams,
which were responsible for the soft-core repulsion.

In this work, we incorporate the dynamics of the N-N interaction by
employing the Generator Coordinate Method (GCM); we derive an
approximate differential equation for the N-N wave function describing
the relative motion of the two nucleons.
This equation contains a local N-N
potential and an effective, coordinate dependent mass. By means of a
Fujiwara transformation, we then define a N-N separation length, x,
from the deformation parameter used previously in the adiabatic
approximation. This allows us to introduce a constant mass and to
rewrite the effective potential in terms of this coordinate x. One of
our objectives is to study the explicit role of the one-gluon exchange
effects on the local N-N potential, included for the first time in
such type of calculations. Another aim is to establish a connection
between our effective deformation parameter and the true internucleon
separation. The latter will enable us to apply our six-quark wave
functions to studies of the quark substructure of light nuclei, as
has been carried out already, for instance, in Ref. \cite{KW95}. The
present numerical results correspond to the (TS)=(10) sector, although
the formalism at hand can easily be extended to other isospin-spin
channels.

\section{The Generator Coordinate Method}

The GCM was introduced in the fifties by Hill and Wheeler \cite{HW53}
to describe collective motion in nuclear systems, such as rotation,
vibration or center of mass motion \cite{GW57,PY57}. Starting from a
many-body wave function $|\,\alpha\,\rangle$ depending on a collective
coordinate $\alpha$ (the deformation parameter of the six-quark system
in our case), a trial wave function is constructed by taking a linear
combination of the states $|\,\alpha\,\rangle$ with some weight
function $\Phi(\alpha)$,
\begin{equation}
|\,\Psi\,\rangle=\int\Phi(\alpha)~|\,\alpha\,\rangle~d\alpha \ ,
\end{equation}
where $\Phi(\alpha)$ is determined through the variational principle
\begin{equation}
\begin{array}{ccccc}
\delta E & = & \displaystyle \frac{\delta}{\delta\Phi^*}\frac{
\langle\,\Psi\,|\,H\,|\,\Psi\,\rangle}{\langle\,\Psi\,|\,\Psi\,
\rangle} & = & 0 \ , \\
\end{array}
\end{equation}
which leads to the Hill-Wheeler integral equation:
\begin{equation}
\int\langle\,\alpha\,|\,H-E\,|\,\alpha'\,\rangle~\Phi(\alpha')~
d\alpha' = 0 \ .
\end{equation}
This is a homogeneous Fredholm-type equation of the first kind,
notoriously unstable numerically. Although some methods exist to make
it stable (such as regularization \cite{WW88}, removal of the zero
normalization eigenmodes \cite{RS80}, Gaussian transform \cite{GT73},
etc.), we prefer to solve a differential equation approximately
equivalent to the Hill-Wheeler equation, both for numerical stability
and to facilitate comparison with analyses based on the Schr\"odinger
equation. In general, $\alpha$ is a multidimensional parameter. It is
at least three-dimensional when correspondence is made to ${\bf r}$.
We here restrict the calculations to the zero-impact parameter case,
which reduces the problem to a one-dimensional one, and leave
consideration of the angles to a later study.

\section{The Hill-Wheeler differential equation}

To derive such a differential equation, it is more convenient to work
with mean and relative deformation parameters, $\beta$ and $\delta$,
defined as
\begin{equation}
\begin{array}{ccc}
   \beta & = & \displaystyle \frac{\alpha +\alpha'}{2} \ ,\\
   \delta & = & \alpha - \alpha' \ .
\end{array}
\end{equation}
Expanding the weight function in a Taylor series around $\delta=0$,
one has:
\begin{eqnarray}
\langle\,\Psi\,|\,H-E\,|\,\Psi\,\rangle~= \int &&d\beta \int d\delta
\left[\Phi^*(\beta)+\frac{\delta}{2}{\Phi^*}'(\beta)+
\frac{\delta^2}{8}{\Phi^*}''(\beta)+\ldots\right]\nonumber\\
&&\langle\,\beta+\frac{\delta}{2}\,|\,H-E\,|\,\beta-\frac{\delta}
{2}\,\rangle\left[\Phi(\beta)-\frac{\delta}{2}\Phi'(\beta)+
\frac{\delta^2}{8}\Phi''(\beta)+\ldots\right] \ .
\end{eqnarray}
It is convenient to introduce the moments:
\begin{eqnarray}
H_n & = & \int d\delta~\langle\,\beta+\frac{\delta}{2}\,|\,H\,|\,
          d\beta-\frac{\delta}{2}\,\rangle~\delta^n \ , \\
N_n & = & \int d\delta~\langle\,\beta+\frac{\delta}{2}\,|\,
          d\beta-\frac{\delta}{2}\,\rangle~\delta^n \ .
\end{eqnarray}
Because $\langle\,\beta+\delta/2\,|\,H-E\,|\,\beta-\delta/2\,\rangle$
is an even function of $\delta$, the odd moments are zero. Supposing,
moreover, that $\langle\,\beta+\delta/2\,|\,H-E\,|\,\beta-\delta/2\,
\rangle$ is a sharply peaked function of $\delta$, one can stop the
expansion at second order in $\delta$. Partial integration and
variation by $\delta \Phi^*$ leads then to the Hill-Wheeler
differential equation:
\begin{equation}
\frac{1}{2}\frac{d}{d\beta}\left((H_2-EN_2)\frac{d\Phi}{d\beta}\right)
 + \left[H_0+\frac{1}{8}\frac{d^2H_2}{d\beta^2}\right]\Phi = E
\left[N_0+\frac{1}{8}\frac{d^2N_2}{d\beta^2}\right]\Phi \ .
\label{eqhw1}
\end{equation}

The introduction of a new function into the hermitian,
\begin{equation}
\tilde{\Phi}(\beta)=\sqrt{\tilde{N}_0(\beta)}~\Phi(\beta) \ ,
\end{equation}
where
\begin{equation}
\tilde{N}_0 = N_0 + \frac{1}{8}\frac{d^2 N_2}{d\beta^2} \ ,
\end{equation}
allows us to transform Eq. (\ref{eqhw1}) into hermitian form:
\begin{equation}
\left[-\frac{d}{d\beta}\frac{1}{2B(\beta)}\frac{d}{d\beta}+V(\beta)
\right]\tilde{\Phi}(\beta) = E\tilde{\Phi}(\beta) \ ,
\label{schrod}
\end{equation}
where $V(\beta)$ is given by:
\begin{equation}
V(\beta) = \frac{\tilde{H}_0}{\tilde{N}_0}+
\frac{1}{2\sqrt{\tilde{N}_0}}\frac{d}{d\beta}
\left((H_2 - EN_2)\frac{d}{d\beta}
\left(\frac{1}{\sqrt{\tilde{N}_0}}\right)\right) \ ,
\label{pot1}
\end{equation}
with
\begin{equation}
\tilde{H}_0 = H_0 + \frac{1}{8}\frac{d^2 H_2}{d\beta^2} \ .
\end{equation}

The term $B(\beta)$ is the effective mass:
\begin{equation}
B(\beta) = -\frac{\tilde{N}_0}{H_2 - EN_2} \ .
\label{mass}
\end{equation}
The total energy E enters the definition of B; its asymptotic form
at threshold is:
\begin{equation}
E=2m_N \ ,
\end{equation}
where $m_N$ is the nucleon mass.
Note that because we didn't incorporate
center of mass corrections the asymptotic value of the potential in
Eq. (\ref{pot1}) is not equal to the experimental value of $2 m_N$.
We have indeed $V(\infty)$ = 2468 MeV when gluons are not included and
$V(\infty)$ = 2240 MeV when they are. In practice, we could obtain a
value closer to the experimental value by subtracting recoil
corrections from the asymptotic energy:
\begin{equation}
m_{N}^{2}=\left(\frac{V(\infty)}{2}\right)^{\!2}
          ~-~\langle\,P^2\,\rangle \ ,
\end{equation}
but we prefer to avoid this step. This simplification does not affect
our conclusions. Following Brink and
Banerjee \cite{BB73}, we replace $E$ in the mass term by:
\begin{equation}
E \rightarrow  \frac{H_0(\beta)}{N_0(\beta)} \ .
\label{repl}
\end{equation}
This approximation is consistent with neglecting higher order
derivatives of the moments in the Hamiltonian.

The moments $H_n$ (n=0, 2) and the corresponding quantities $B(\beta)$
and $V(\beta)$ have been calculated for three distinct cases:
\begin{equation}
\begin{array}{cccc}
(a)& H &=& H_{1}^{bag} + H_{OGE} \ , \\
(b)& H &=& H_{1}^{bag} + H_{OGE}^{mag} \ , \\
(c)& H &=& H_{1}^{bag}  \ ,
\end{array}
\label{case}
\end{equation}
where $H_{1}^{bag}$, $H_{OGE}^{mag}$ and $H_{OGE}$ are, respectively,
the non-gluonic one-body term of the Hamiltonian, the color-magnetic
and the full one-gluon exchange contribution; they are given
explicitly in Ref. \cite{KW94}. In case (c), the one-gluon
exchange was left out
altogether. This is in the spirit of an earlier investigation where
the Friedberg-Lee soliton model was applied to N-N scattering
without considering gluonic degrees of freedom \cite{Sc86}.
In case (b), the
color-magnetic hyperfine interaction was accounted for, and in case
(a) the full color-magnetic and color-electrostatic OGE was included.
The reason to distinguish between cases (a) and (b) is that in the
literature it was claimed that the color-magnetic part of the OGE
itself is responsible for the repulsive core of the N-N interaction
\cite{barn}. We shall return to this point at the end of Section V.

The plot of $B$ as a function of $\beta$ is given in Fig. 1 for the
three cases (a), (b) and (c). $B$ converges towards a constant
value $\mu$, which can be calculated from considering two
well-separated non-interacting three-quark bags:
\begin{mathletters}
\begin{eqnarray}
\mu &=& 763.6 \mbox{MeV}\hspace{2cm} \mbox{in cases (a) and (b)} \ ,
\label{mue1}
\\
\mu &=& 502.3 \mbox{MeV}\hspace{2cm} \mbox{in case (c)} \ .
\label{mue2}
\end{eqnarray}
\end{mathletters}
We would expect $\mu$ to be equal to the reduced mass, $m_N$/2. The
discrepancy between the values of $\mu$ and $m_N$/2 -- which is
especially drastic if the OGE is included, i.e., in cases (a) and (b)
-- is related to the well-known Peierls-Yoccoz disease
\cite{PY57,RS80}.

\figa{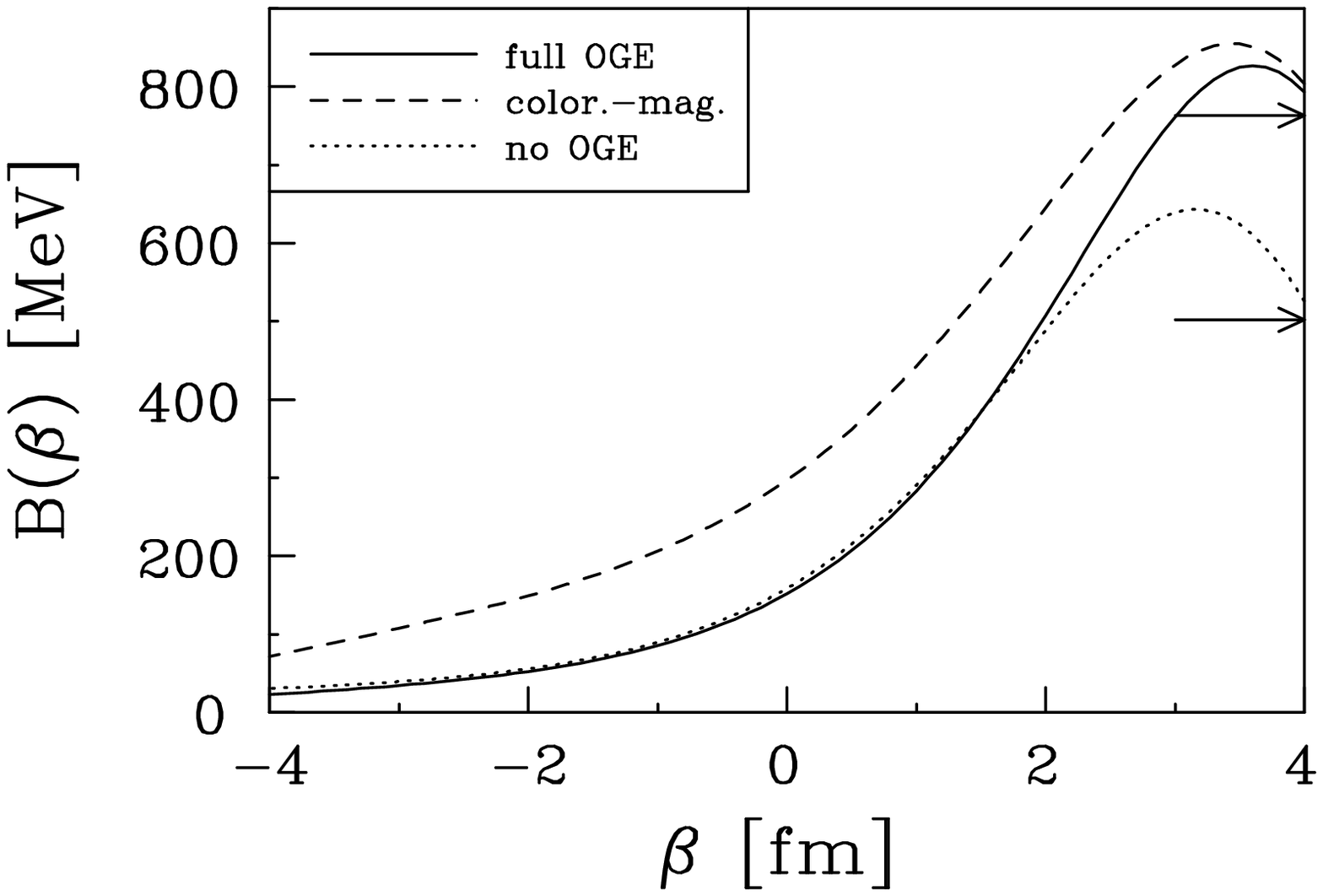}{ht}{The effective mass, $B(\beta)$ of Eq.
(\protect\ref{mass}), as a function of the deformation parameter
$\beta$; the solid, dashed and dotted lines correspond,
respectively, to the cases (a), (b) and (c) introduced in Eq.
(\protect\ref{case}). The asymptotic values of Eqs.
(\protect\ref{mue1}) and (\protect\ref{mue2}) are indicated by
the arrows.}

\section{The Fujiwara transformation}

The dependence of the effective mass on $\beta$ prevents us from
directly interpreting the potential in Eq. (\ref{pot1}) as an ordinary
N-N potential. Moreover, $\beta$ doesn't correspond to the true N-N
separation distance (except for large positive $\beta$ when the two
nucleons are well separated). Therefore, we wish to transform Eq.
(\ref{schrod}) into a Schr\"{o}dinger-like equation with a constant,
coordinate independent mass term. For this purpose, one can use a
Fujiwara transformation \cite{Fu59,Wi89}, which relates the generator
coordinate $\beta$ to an effective N-N separation length:
\begin{equation}
x(\beta)=-\int^{\infty}_{\beta}\left[\sqrt{\frac{B(\beta')}{\mu}} -1
\right] d\beta' + \beta \ .
\label{fuji}
\end{equation}

If one now redefines the weight function in Eq. (\ref{schrod}) as
\begin{equation}
\tilde{\Phi}(\beta) = \root {\scriptstyle 4}
\of {\frac{B(\beta)}{\mu}}~ \psi(x) \ ,
\end{equation}
Eq. (\ref{schrod}) transforms into the familiar form
\begin{equation}
\left[-\frac{1}{2\mu}\frac{d^2}{dx^2}+V+V_F\right]\psi(x)=E\psi(x) \ ,
\end{equation}
with V given by Eq. (\ref{pot1}) and
\begin{equation}
V_F(\beta) = \frac{7}{32B^3}\left(\frac{dB}{d\beta}\right)^{\!2}
            -\frac{1}{8B^2}\frac{d^2B}{d\beta^2} \ .
\label{potf}
\end{equation}

\figa{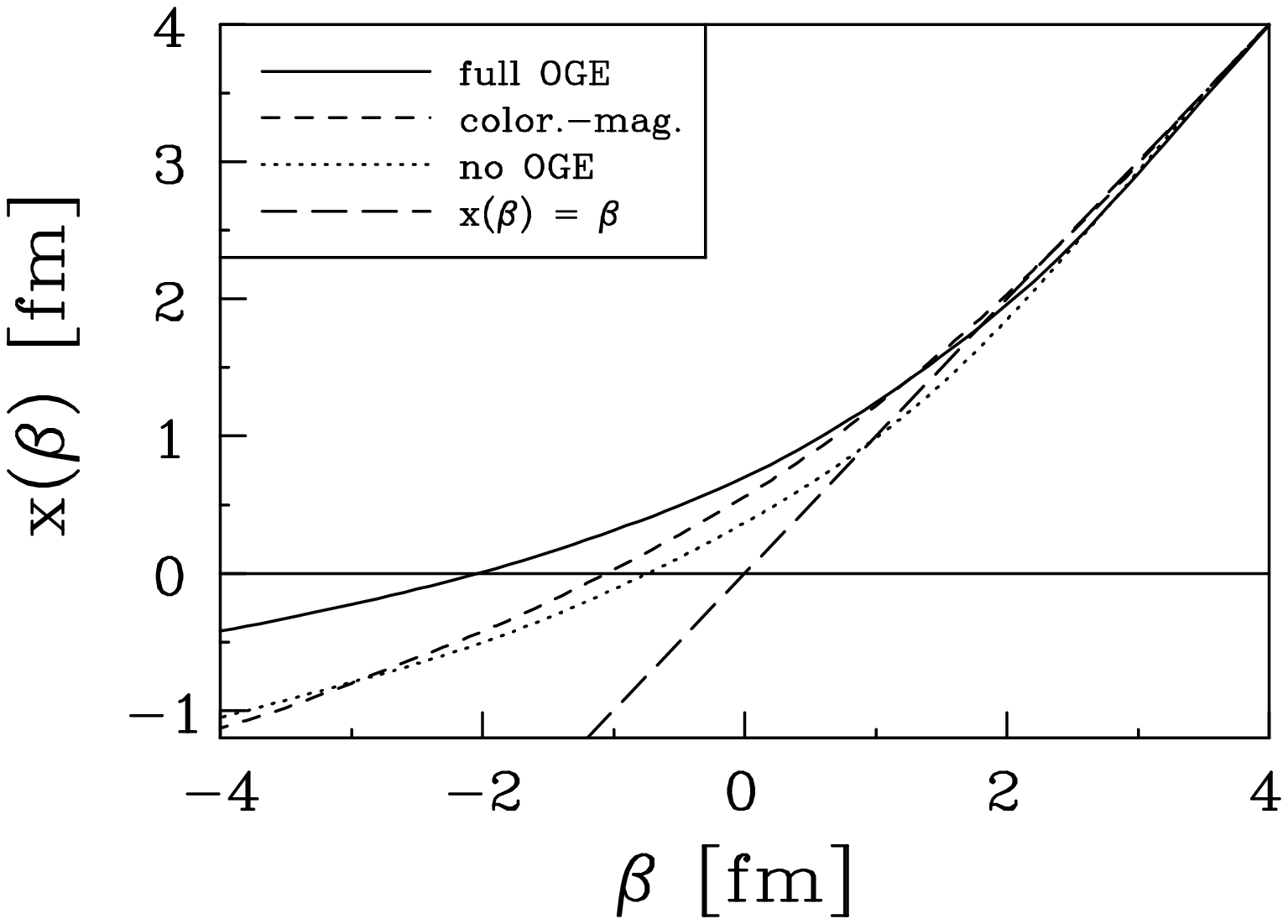}{bht}{The Fujiwara coordinate, $x(\beta)$ of Eq.
(\protect\ref{fuji}), as a function of the deformation parameter
$\beta$. The long-dashed line shows the asymptotic solution,
$x(\beta)=\beta$, and the remaining labeling is the same as in
Fig. 1.}

Figure 2 displays the explicit relationship between $x$ and $\beta$,
as obtained from Eq. (\ref{fuji}). As expected, the deformation
parameter $\beta$ converges asymptotically towards the effective
internucleon separation $x$. The correspondence
$\beta \leftrightarrow x$ should be
very useful in discussions of the quark substructure of nuclei or
nuclear matter using Schr\"odinger-based many-nucleon calculations
and employing our six-quark wave functions.

\section{Results for the effective N-N potential}

We now wish to present detailed results for:
\begin{equation}
V_{loc}(\beta) = V(\beta) + V_F(\beta) - V(\infty) \ ,
\label{sum}
\end{equation}
where $V(\beta)$ and $V_F(\beta)$ are given in Eqs. (\ref{pot1}) and
(\ref{potf}). The value of $V(\infty)$ corresponds to the asymptotic
value of $\tilde H_0/\tilde N_0$ calculated from two well-separated
non-interacting three-quark bags, and it is given in Section III.
This asymptotic value is the same in cases (a) and (b) because the
color-electrostatic mutual and self-energy terms cancel exactly due to
color neutrality when the two nucleons are well separated.

\figb{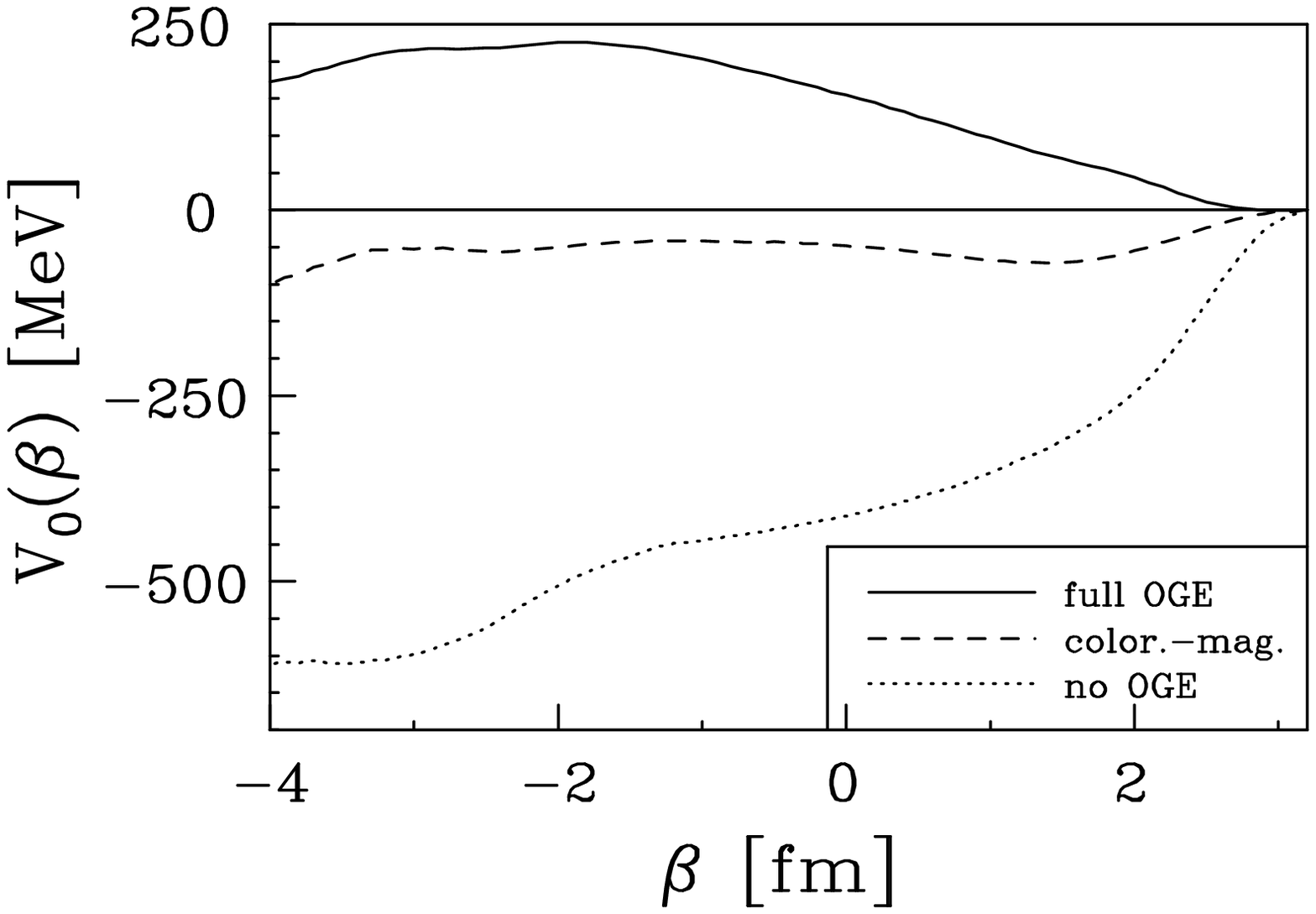}{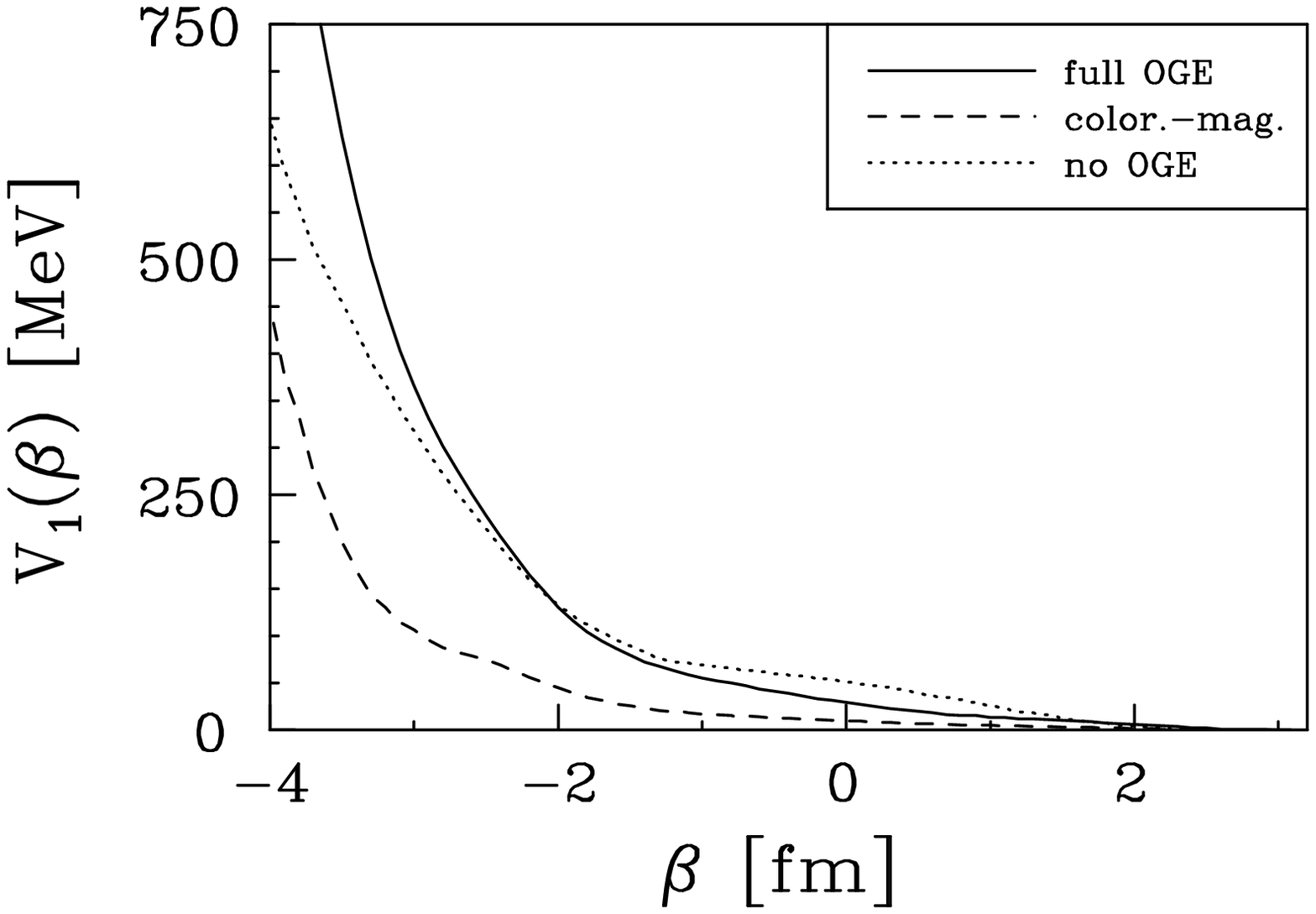}{p}{The two contributions to the local potential,
$V_0(\beta)$ of Eq. (\protect\ref{v1}) and $V_1(\beta)$ of Eq.
(\protect\ref{v2}), as functions of the mean generator coordinate
$\beta$. The solid, dashed and dotted lines correspond
to the cases (a), (b) and (c) introduced in Eq.
(\protect\ref{case}).}

\figb{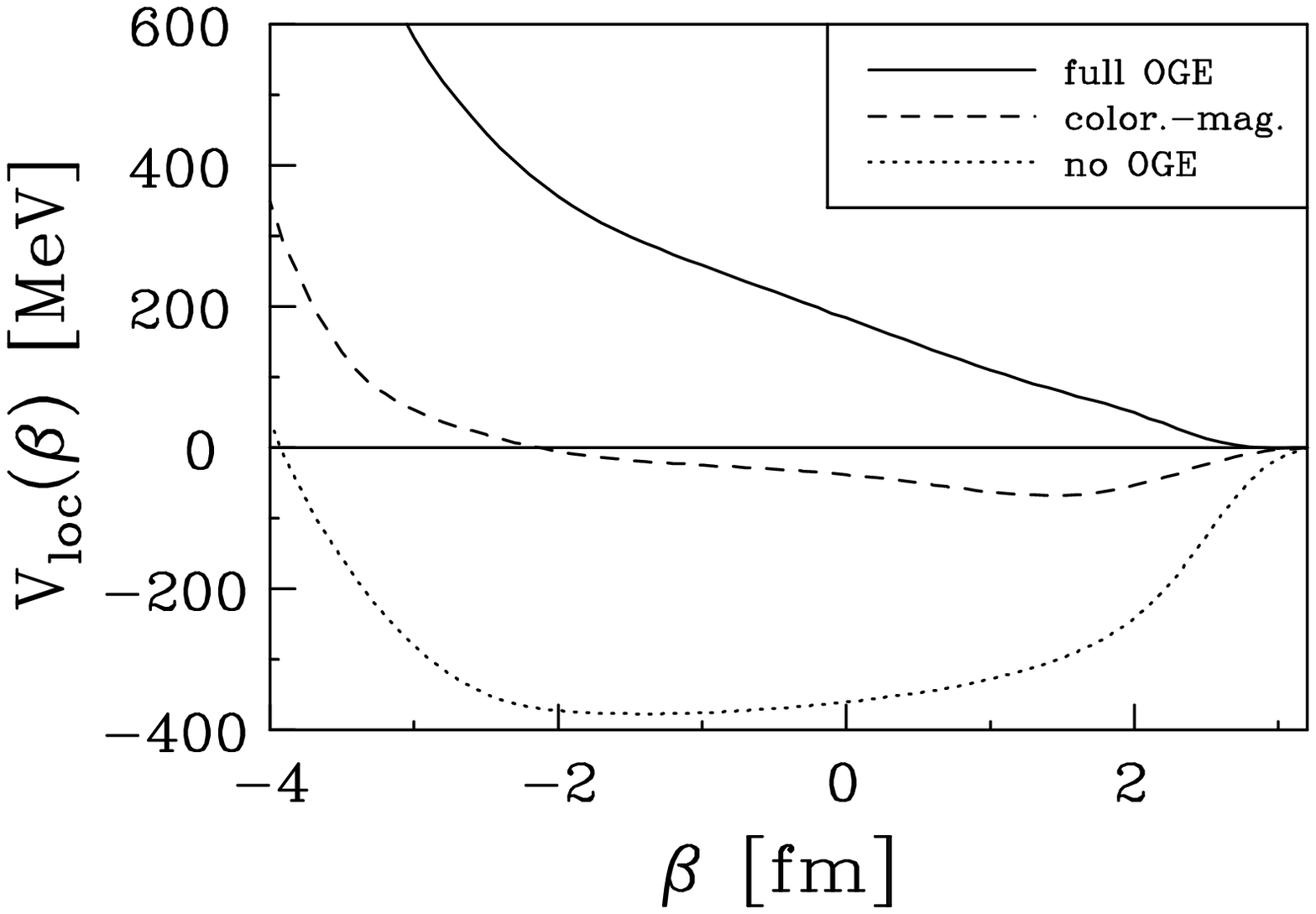}{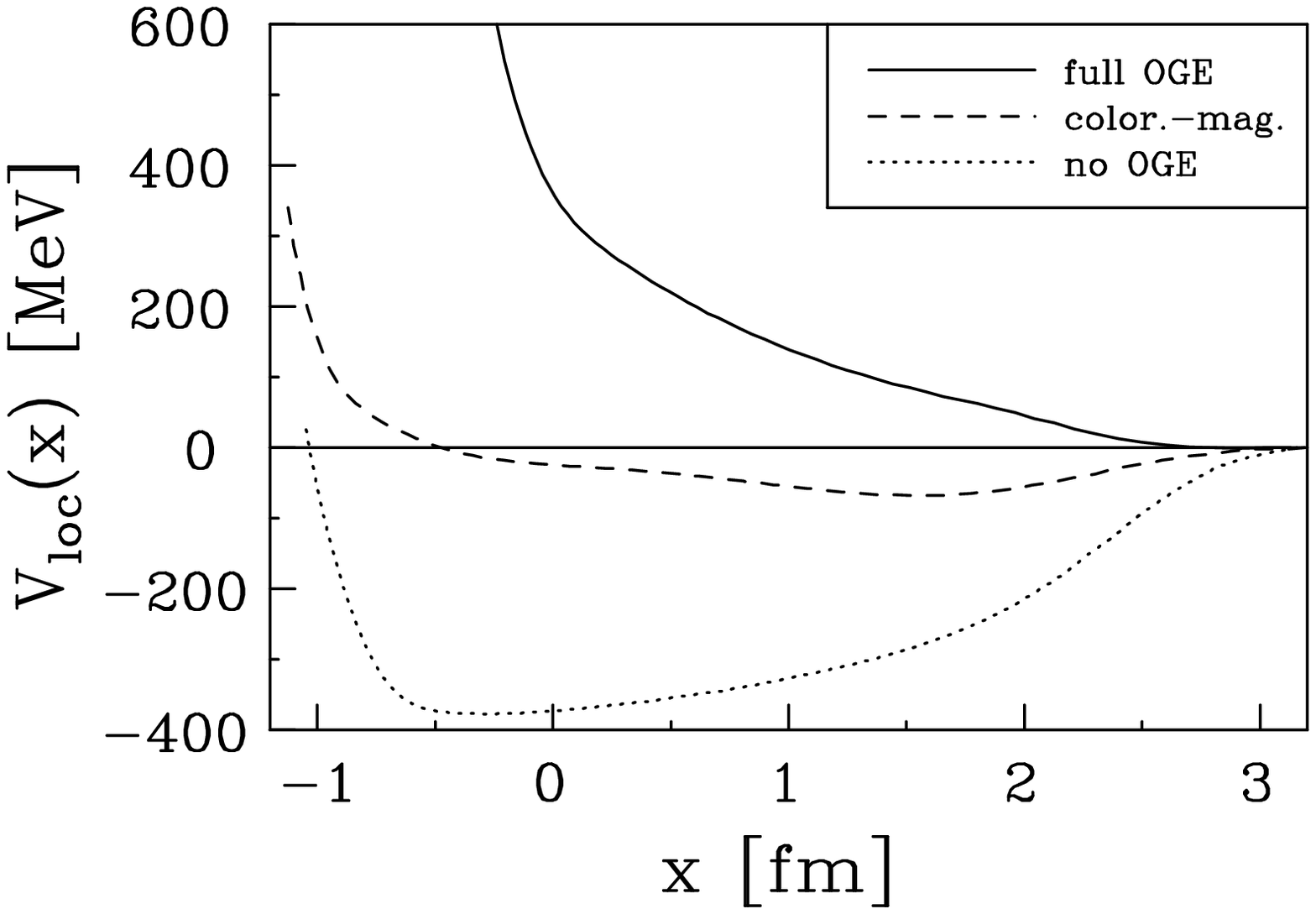}{p}{The non-adiabatic, local potential, $V_{loc}$
of Eq. (\protect\ref{vloc}), as a function of the deformation
parameter $\beta$ and the Fujiwara coordinate $x$, respectively. The
labeling is the same as in Fig. 3.}

It is convenient to rewrite $V_{loc}(\beta)$ in the following form:
\begin{equation}
V_{loc}(\beta)=V_0(\beta)+V_1(\beta) \ ,
\label{vloc}
\end{equation}
with:
\begin{eqnarray}
V_0(\beta) &=& \frac{\tilde{H}_0}{\tilde{N}_0}-V(\infty) \ ,
\label{v1}\\
V_1(\beta) &=& \frac{1}{4B}\left[
  \frac{d^2\ln\tilde{N}_0}{d\beta^2}
 +\frac{1}{2}\left(\frac{d\ln\tilde{N}_0}{d\beta}\right)^{\!2}
 -\frac{d\ln\tilde{N}_0}{d\beta}\frac{d\ln B}{d\beta}\right]
\nonumber\\
& &~~~+~\frac{1}{8B}\left[
  \frac{3}{4}\left(\frac{d\ln B}{d\beta}\right)^{\!2}
  -\frac{d^2\ln B}{d\beta^2}\right] \ .
\label{v2}
\end{eqnarray}
In order to calculate these derivatives, $\ln{B}$ and
$\ln{\tilde{N}_0}$ were fitted to polynomials.
The two contributions $V_i(\beta)$ (i=0, 1) to $V_{loc}(\beta)$
are plotted in Fig. 3 as functions of the deformation
parameter $\beta$ for the three cases outlined previously. Fig. 4
shows $V_{loc}$ as a function of $\beta$ and of the
Fujiwara coordinate $x$, respectively. Note that Eq. (\ref{v2}) was
obtained from Eq. (\ref{pot1}) by replacing $H_2 - E N_2$ with
$-\tilde N_0/B$, as indicated in Eq. (\ref{mass}).

The shape of $V_0(\beta)$ is quite similar to the adiabatic potentials
displayed in Fig. 10 of Ref. \cite{KW94}, both for the ``full OGE" and
 ``no-OGE" cases. This tends to confirm our
assumption that the matrix elements $\langle\,\beta+\delta/2\,|
\,H-E\,|\,\beta-\delta/2\,\rangle$ are rather sharply peaked around
$\delta=0$. The term $V_1(\beta)$ corresponds to the contribution of
non-adiabaticity. It grows important only for $\beta \lesssim -2$ fm,
and yields in all cases a repulsion due to the dynamics. This is
according to our expectation and in agreement with Ref.\cite{Sc86}.
Note that
in cases (b) and especially (c), we also obtain an intermediate range
attraction in $V_{loc}$.
The fact that our N-N potential extends to negative $x$ should not be
taken too literally. It simply reflects inadequacies in the
relationship between the deformation parameter $\beta$ and the N-N
separation length $x$, which are connected to the Peierls-Yoccoz
disease mentioned earlier.

We recall that one of the main objectives of this and our previous
study \cite{KW94} was to incorporate explicitly one-gluon exchange
effects, in contrast to Ref.\cite{Sc86} where they were neglected.
Comparing, for instance, cases (a) and (c), one can see that the OGE
reinforces the repulsive core considerably. The existence of a
repulsive core in all three cases makes us to attribute it to
dynamics rather than to the color-magnetic interaction (case (b)), as
was inferred in Ref.\cite{barn}.

\section{Conclusions}

In this investigation, we found that the dynamics are manifestly
responsible for the hard-core repulsion of the short-range part of the
N-N interaction, and we observed that we could obtain both short-range
repulsion and some intermediate range attraction if the entire
one-gluon exchange or at least its electrostatic part were neglected.

In the results containing the full OGE effects the lack of
attraction
is due to the omission of explicit meson exchanges. Then, to reproduce
the experimental phase shifts or other two-body data one necessitates
to attach a local OBE potential beyond a certain internuclear distance
\cite{Oka83}. To obtain this potential in the framework of our model
we could consider extending our calculations by either including
quantum surface fluctuations and/or introducing configurations of the
form $q^7\bar{q}$ in addition to the $q^6$ states. This would be a
rather cumbersome procedure within the present model. The most
convenient would be to either allow mesonic degrees of freedom and
to consider, e.g., an explicit pion exchange between the individual
quarks \cite{myhr} or to simply choose a phenomenological potential.

Another important result of this work is the evaluation of the
relationship between the deformation parameter $\beta$ and the
effective N-N separation length $x$ through the Fujiwara
transformation. This correspondence is very useful for applications
of our model to the description of phenomena involving the quark
substructure of light nuclei. It furthermore allows us to relate
many-body correlation functions or N-N wave functions given in the
literature to the GCM formalism presented here.

An attractive way to confirm our results would be to solve directly
the Hill-Wheeler integral equation in order to obtain phase shifts.
Projection on good angular momentum states should also improve our
calculations.

\acknowledgments{This work was supported in part by the MINERVA
Foundation of the Federal Republic of Germany, and by the U.S.
Department of Energy.}


\end{document}